\documentclass[12pt,preprint]{aastex}

\newdimen\minuswidth    
\setbox0=\hbox{$-$}
\minuswidth=\wd0
\catcode`@=\active
\def@{\kern\minuswidth}
\newdimen\digitwidth    
\setbox0=\hbox{\rm0}
\digitwidth=\wd0
\catcode`!=\active
\def!{\kern\digitwidth}

\begin{document} 
  
\newcommand{\gras}[1]{\mbox{\boldmath $#1$}}
\newcommand{\xmm}{XMM-{\em Newton}}
\newcommand{\chandra}{{\em Chandra}}


\title{Stellar and gaseous abundances in M82\altaffilmark{1}}
\altaffiltext{1}{Based on observations with the Near Infrared Camera 
Spectrometer 
(NICS) mounted at the Telescopio Nazionale Galileo (TNG), La Palma, Spain and 
archival \xmm\ and \chandra\ data. 
}

\author{L. Origlia} 
\affil{INAF -- Osservatorio Astronomico di Bologna,
Via Ranzani 1, I--40127 Bologna, Italy,\\
origlia@bo.astro.it}
\author{P. Ranalli} 
\affil{Dipartimento di Astronomia Universit\`a 
di Bologna, Via Ranzani 1, I--40127 Bologna, Italy,
ranalli@bo.astro.it}
\author{A. Comastri} 
\affil{INAF -- Osservatorio Astronomico di Bologna,
Via Ranzani 1, I--40127 Bologna, Italy,\\
comastri@bo.astro.it}
\and
\author{R. Maiolino}
\affil{INAF -- Osservatorio Astrofisico di Arcetri,
Largo E. Fermi 5, I--50125 Firenze, Italy,
maiolino@arcetri.astro.it}

  
\begin{abstract}
The near infrared (IR) absorption spectra of starburst galaxies 
show several atomic and molecular lines from red supergiants  
which can be used to infer reliable stellar
abundances.  The metals locked in stars give a picture of the galaxy
metallicity prior to the last burst of star formation.
The enrichment of the new generation of stars born in the last burst
can be traced by measuring the hot gas in the X-rays. 
For the first time detailed stellar abundances in the nuclear region of the starburst  
galaxy M82 have been obtained. 
They are compared with those of the hot gas
as derived from an accurate re-analysis of the 
\xmm\ and \chandra\ nuclear X-ray spectra. 
The cool stars and the hot gas
suggest [Fe/H]$=-0.35\pm0.2$~dex, and an overall [Si,Mg/Fe]
enhancement by $\simeq $0.4 and 0.5~dex, respectively.
This is consistent with a 
major chemical enrichment by SNe~II explosions in recursive bursts on
short timescales.  
Oxygen is more puzzling to interpret since 
it is enhanced by $\simeq $0.3~dex
in stars and depleted by $\simeq $0.2~dex in the hot gas. None of the 
standard enrichment scenarios can fully explain such a behavior when compared with 
the other $\alpha$-elements. 

\end{abstract}


\section{Introduction} 
\label{intro}

The signature of the star formation (SF) history of a galaxy is
imprinted in the abundance patterns of its stars and gas.  Determining
the abundance of key elements released in the interstellar medium
(ISM) by stars with different mass progenitors and hence on different
time scales, will thus have a strong astrophysical impact in drawing
the global picture of galaxy formation and evolution
\citep{mwi97}.   

So far, the metallicities of distant starburst (SB) galaxies have been
mainly derived by measuring the nebular lines associated to their
giant HII regions \citep{sck94,coz99,hek00}. 
Direct determination of O and N abundances can be obtained once electron
temperature (T$_e$) is well established \citep{mca84}.  
At high metallicities T$_e$ decreases due to strong lines cooling and 
its major diagnostic (the weak [OIII] $\lambda$4636) is no
longer observable. 
Several alternative methods based on other strong lines were proposed
in the past years \citep{pag79,de86,mcg91} to derive T$_e$,
but all of them rely on a large number of physical parameters and
model assumptions for the geometry and the nature of the ionizing
populations (c.f. e.g \citet{sta01} for a review).  
Depending on the adopted calibrations, quite different abundance sets 
can reproduce the observed line ratios.

However, abundances in SB galaxies can be also measured  
from absorption stellar features in the near IR  
and/or emission lines in the hot X-ray gas.
Stellar abundances can be poorly constrained
from optical spectra, since the nebular emission
strongly dilutes the absorption lines and dust can heavily obscure
the central regions where most of the burst activity is concentrated.
On the contrary, in the near IR the stellar continuum due to red supergiants (RSGs) 
usually largely dominates over the possible gas and dust emission
\citep{oo98,oo00}.
The X-ray emission from SB galaxies at E$>$2 keV is mainly due to 
high mass X-ray binaries with power-law spectra, while at lower energies 
hot plasma mainly heated by SN explosions \citep{dah98} dominates 
and the resulting spectrum is a superposition of thermal continuum and
emission lines from atomic recombinations \citep{fab89,pr02}.

Determining the abundances of SB 
galaxies is not only relevant to constrain their SF  
history, but also offers the unique chance of directly
witnessing the enrichment of the ISM 
\citep[see e.g.][ for a review]{mc94}.
Metals locked into stars give a picture
of the enrichment just prior to the last burst of SF,
while the hot gas heated by SNe~II explosions 
should trace the enrichment by the new generation
of stars. 
Nebular metallicities from the cold gas should potentially trace the stellar ones
if the cooling
and mixing timescales are slow (up to 1 Gyr), as suggested by some models
\citep[see e.g.][]{tt96}
since the gas enriched by the ongoing burst is still too hot to be
detected in the optical lines.
On the contrary, if rapid (a few Myr) cooling and mixing occur
as suggested by other models \citep[see e.g.][]{rmde01},
the nebular abundances could trace the gas enriched by the
new population of stars (similarly to X-rays).
Hence, important constraints on the cooling and mixing
timescales of the gas can be provided
by comparing the metallicity inferred from stars and hot plasma
with those of the cold gas.                                  

Sect.~\ref{lit} briefly reviews the nebular abundance estimates of M82, 
as published so far in the literature. Sect.~\ref{ir} shows our near 
IR spectra of the nuclear region of M82
and the derived stellar abundances. Sect.~\ref{xray} presents a re-analysis of the 
\xmm\ and \chandra\ nuclear X-ray spectra and the hot gas--phase abundances. 
Sect.~\ref{disc} summarized the overall chemical enrichment scenario in M82 as traced by 
the stellar, cold and hot gas-phase abundances, while in Sect.~\ref{conc} we draw our 
conclusions.
 
\section{Nebular abundances of M82 from the literature}
\label{lit}

M82 is a prototype of SB galaxies \citep{rie80}, experiencing a major  
SF formation episode in its nuclear regions, with strong super-wind 
and SN activity \citep{wil99}.
Despite the many multi-wavelength studies 
\citep[see e.g.][and references therein]{gak96}, 
only few measurements of nebular abundances, mainly derived 
from mid/far IR forbidden lines, have 
been published so far.
From the analysis of the [SiII] 35$\mu$m line in the central 300~pc in radius 
at the distance of M82 (i.e. d$\simeq $3.6~Mpc),
\citet{l96} suggest an indicative  
$\rm [Si/H]\simeq -0.4$~dex relative to the Solar value
of \citet{grev98}.

More recently, \citet{fs01}, by analyzing the nuclear spectra of M82 taken 
with the Short Wavelength Spectrometer (SWS)
onboard the Infrared Space Observatory (ISO) 
within an aperture corresponding to $100\times200$~pc on the sky, 
give cold gas-phase abundances of 
[Ne/H]=+0.1, [Ar/H]=+0.3 and [S/H]=--0.7 dex 
relative to the Solar values
of \citet{grev98}, with an overall uncertainty between 
0.2 and 0.3 dex. 

By taking the line ratios listed by \citet{acsj79} which refer to the central 
$\simeq $100 pc 
in radius of M82 \citep{ps70}, and the empirical calibrations of 
\citet{mcg91} which give N and Fe abundances as a function of [O/H]                   
(see e.g. their Figure~11, equations~9 and 10a), one 
can also obtain rough estimates of 
[O/H]=+0.1, [N/H]=0.0 and [Fe/H]=--0.3 dex relative to the Solar values 
of \citet{grev98}, with an overall uncertainty of $\pm$0.3~dex. 

The first attempt to measure hot gas-phase abundances in M82 
has been made quite recently by \citet{pta97}
and slightly revised by \citet{ume02},
by analyzing ASCA X-ray (0.4-10.0~keV) spectra.
Fe and O abundances as low as 1/20 Solar and 
significantly higher (by a factor between 3 and 10) Si, S, Mg, Ca
ones have been obtained, also consistent with BeppoSAX results
\citep{cappi99}.

More recently, \citet{rs02} by analyzing the spectra obtained with the 
Reflection Grating Spectrometer (RGS) onboard \xmm\
found near-Solar Fe and O and super-Solar 
(between a factor of 2 and 5) Mg, Si, Ne, N abundances.
While the abundance ratios are in reasonable agreement 
within $\pm$0.3~dex with \citet{pta97}, there is one order magnitude 
difference between the zero-point of the two calibrations.

\section{IR spectra and stellar abundances}
\label{ir}

Near-IR spectroscopy is a fundamental tool to obtain accurate 
abundances of key elements like Fe, C, O 
and other metals (e.g. Si, Mg, Ca, Na, Al, Ti) 
in cool stars ($\rm T_{eff}\lesssim 5000$~K).
Several atomic and molecular lines are strong,
not affected by severe blending, hence also measurable at low-medium resolution
\citep{kh86,o93,wh97,joy98,fro01},
making them powerful abundance tracers not only in stars but also in more 
distant stellar clusters and galaxies and in a wide range of metallicities and ages
\citep{ori97,oo98}.
Abundance analysis from integrated spectra of galaxies 
requires full spectral synthesis techniques to properly account for line blending 
and population synthesis to define the 
dominant contribution to the stellar luminosity.

In SB galaxies the stellar IR continuum usually dominates
over the possible gas and dust emission. 
This represents a major, conceptual simplification in population and
spectral synthesis techniques, making possible and easier the interpretation
of integrated spectra from distant stellar clusters and galaxies. 

\begin{figure}[htbp] 
\plotone{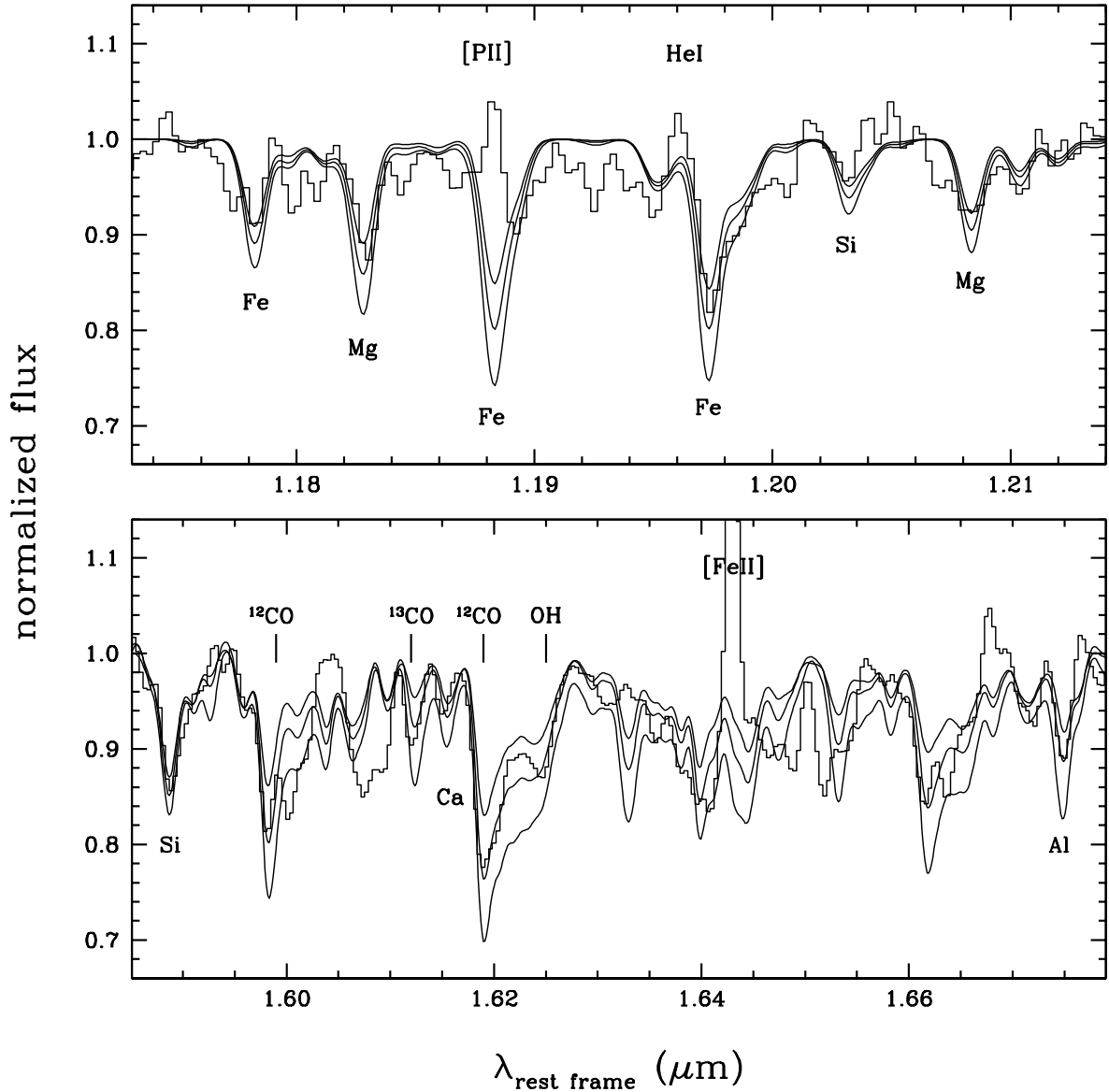} 
\caption {Near-IR spectra of the nuclear region of M82.
Observed spectra: histograms; synthetic stellar best-fit solution and
two other models with $\pm$0.3 dex abundance variations: solid lines.
A few major stellar and nebular lines are also marked. 
\label{IR}}
\end{figure}

Since the near IR continuum of SB galaxies is dominated 
by luminous RSGs \citep[see e.g.][for a review]{oo00},
their integrated spectrum can be modeled with
an equivalent, average star, whose stellar parameters 
(temperature T$_{eff}$, gravity log{\it g} and microturbulence velocity $\xi$)
mainly depend on the stellar age and metallicity.
Both observations and evolutionary models 
\citep[see e.g.][ and reference therein]{k99,o99,ori03}
suggest that RSGs of ages between $\simeq $6 and 100 Myr and 
metallicities between 1/10 and Solar 
are characterized by low gravities 
(log~g$<$1.0), low temperatures ($\le$ 4000~K) 
and relatively high microturbulence velocity ($\xi\ge$3 km/s).  
A variation in the adopted stellar parameters for the average RSG population
by $\Delta $T$_{\rm eff}$$\pm$200~K, $\Delta $log~g=$\pm$0.5 and
$\Delta \xi$$\mp$0.5~~km~s$^{-1}$
implies a $\le \pm 0.1$~dex change in the abundances estimated from atomic lines,
and $\le \pm 0.2$~dex in those estimated from the molecular lines, which are  
more sensitive to stellar parameters.

1.0--1.8$\mu$ longslit spectra (see Fig.~\ref{IR}) 
at R=2,500 of the nuclear region of M82
were obtained with the Near IR Camera Spectrometer (NICS) 
mounted at the Italian Telescopio Nazionale Galileo (TNG) on December 2002.  
The spectra were sky-subtracted, flat-fielded and
corrected for atmospheric absorption using an
O-star spectrum as reference.
They were wavelength calibrated by using a Ne-Ar lamp and the 
monodimensional spectra were extracted by summing over 
the central $0.5\arcsec\times 3.0\arcsec$, corresponding to 
 an aperture projected on the source of $9\times 52$~pc at the distance of M82.  
The spectra have been normalized to the continuum, 
which was determined applying a low-pass smoothing filter to each spectrum.    
Total integration times of 72 and 32 min in the J and H bands, respectively, 
were used, providing a final signal to noise ratio $\ge$40.

By measuring the absorption line broadening a stellar
velocity dispersion $\sigma \simeq 105 \pm$20 km/s has been
derived, in perfect agreement with the estimate by 
\citet{glt93}
and in the typical range of values measured in other
massive SB galaxies \citep{oli99}.

\begin{figure}[htbp] 
\plotone{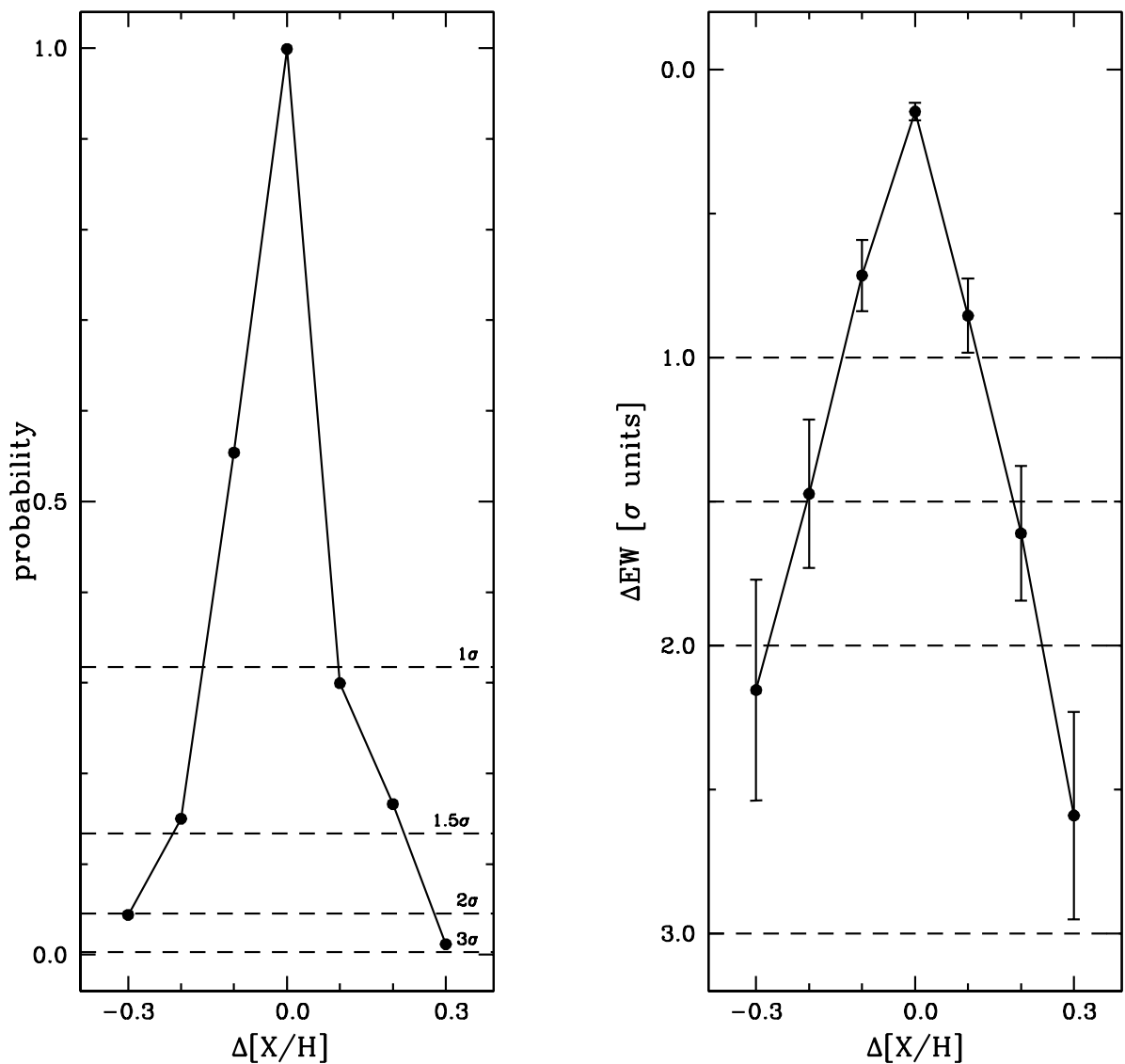} 
\caption{{\it Left panel}:  probability of a random realization
of our best-fit full spectral synthesis solution with varying the elemental abundances
$\Delta[X/H]$ of $\pm$0.2 and 0.3 dex with respect to
the best-fit (see Sect.~\ref{ir}). 
{\it Right panel}: average
difference between the model and the observed equivalent widths of a few 
selected lines (see Sect.~\ref{ir} and Fig.~\ref{IR}).  
\label{IRtest}}
\end{figure}

A grid of synthetic spectra of red supergiant stars
for different input atmospheric parameters and abundances
have been computed, using an updated \citep{o02,o03}
version of the code described in \citet{o93}.
Briefly, the code uses the LTE approximation and is based
on the molecular blanketed model atmospheres of
\citet{jbk80}. It includes several thousands of near IR atomic lines and molecular 
roto-vibrational transitions due to CO, OH and CN.
Three main compilations of
atomic oscillator strengths are used, namely
the Kurucz's database 
(c.f. {\it http://cfa-www.harward.edu/amdata/ampdata/kurucz23/sekur.html}), 
and those published by \citet{bg73} and \citet{mb99}.
 
Our code provides full spectral synthesis over the 1-2.5 $\mu$m range.
Given the high degree of line blending, our abundance estimates
are mainly obtained by best-fitting the full
observed spectrum and by measuring the equivalent widths
of a few selected features (cf. Fig.~\ref{IR}), dominated by 
a specific chemical element, as a further cross-check. 
The equivalent widths have been measured by performing a Gaussian fit 
with $\sigma$ equal to the measured stellar velocity dispersion, 
typical values ranging between 0.5 and 3 \AA, 
with a conservative error of $\pm$200~m\AA\ to also account for a $\pm$2\% 
uncertainty in the continuum positioning.

\begin{deluxetable}{lcccc}
\label{ab}
\tablewidth{10.2cm}
\tablecaption{Element abundances in Solar units$^a$ as derived from our analysis of the IR and X-ray spectra.}
\tablehead{
\colhead{}&
\colhead{stellar$^b$}&
\colhead{}&
\colhead{}&
\colhead{hot gas-phase$^c$}
}
\startdata
$\rm [Fe/H]$ &  $-0.34\pm0.20$ &&& $-0.37\pm0.11$\\
~Fe/Fe$_{\odot}$ &$0.46^{+0.26}_{-0.17}$  &&& $0.43^{+0.12}_{-0.08}$ \\
\hline
$\rm [O/H]$  &  $+0.00\pm0.16$ &&& $-0.58\pm0.19$\\
~O/O$_{\odot}$   &$1.00^{+0.46}_{-0.32}$  &&& $0.26^{+0.15}_{-0.09}$\\
\hline
$\rm [Ca/H]$ &  $+0.05\pm0.28$ &&& --              \\
~Ca/Ca$_{\odot}$ &$1.11^{+1.01}_{-0.52}$  &&& --\\
\hline
$\rm [Mg/H]$ &  $+0.02\pm0.15$ &&& $+0.13\pm0.09$  \\
~Mg/Mg$_{\odot}$~~~~~ &$1.06^{+0.44}_{-0.31}$  &&& $1.36^{+0.32}_{-0.26}$\\
\hline
$\rm [Si/H]$ &  $+0.04\pm0.28$ &&& $+0.17\pm0.08$  \\
~Si/Si$_{\odot}$ &$1.09^{+0.99}_{-0.52}$  &&& $1.49^{+0.32}_{-0.26}$\\
\hline
$\rm [Al/H]$ &  $+0.23\pm0.20$ &&& --\\
~Al/Al$_{\odot}$ &$1.69^{+0.97}_{-0.62}$  &&& --\\
\hline
$\rm [C/H]$  &  $-0.60\pm0.10$ &&& --\\
~C/C$_{\odot}$   &$0.25^{+0.06}_{-0.05}$  &&& --\\
\hline
$\rm [Ne/H]$ &  --             &&& $-0.35\pm0.14$\\     
~Ne/Ne$_{\odot}$ &--                      &&& $0.45^{+0.17}_{-0.12}$\\
\hline
$\rm [S/H]$  &  --             &&& $+0.15\pm0.13$\\    
~S/S$_{\odot}$   &--                      &&& $1.42^{+0.48}_{-0.40}$\\              
\enddata
\tablecomments{
$^a$ Solar values from \citet{grev98}.\\
$^b$ Stellar abundances from TNG/NICS near IR absorption spectra of red supergiants 
(see Fig.~\ref{IR} and Sect.~\ref{ir}). \\
$^c$ Hot gas-phase abundances from XMM/RGS and {\it pn} spectra 
(see Figs.~\ref{XR},\ref{fig-abun} and Sect.~\ref{xray}).}  
\end{deluxetable}  

By fitting the full observed IR spectrum and by measuring the equivalent
widths of selected lines,
we obtained the following best-fit stellar parameters and abundance 
patterns for M82:
T$_{eff}$=4000, log{\it g}=0.5, $\xi$=3, 
[Fe/H]=--0.34; [O/Fe]=+0.34, [$<$Si,Mg,Ca$>$/Fe]=+0.38; 
[Al/Fe]=+0.5; [C/Fe]=--0.26;
$^{12}$C/$^{13}$C$<$10.
Table~\ref{ab} lists the derived abundances and their associated random errors 
at 90\% confidence. Reference Solar abundances are from \citet{grev98}.

Synthetic spectra with lower element abundances are 
{\em systematically} shallower than the best-fit
solution, while the opposite occurs
when higher abundances are adopted.
In order to check the statistical significance of our best-fit solution, 
as a function of merit we adopt
the difference between the model and the observed spectrum (hereafter $\delta$).
In order to quantify systematic discrepancies, this parameter is
more powerful than the classical $\chi ^2$ test, which is instead
equally sensitive to {\em random} and {\em systematic} scatters 
\citep[see][for more details]{o03}.

Since $\delta$ is expected to follow a Gaussian distribution,
we compute $\overline{\delta}$ and the corresponding standard deviation
($\sigma$) for the best-fit solution and 6 {\it test models}
with abundance variations $\Delta[X/H]=\pm$0.1, 0.2 and 0.3 dex
with respect to the best-fit.
We then extract 10000 random subsamples from each
{\it test model} (assuming a Gaussian distribution)
and we compute the probability $P$
that a random realization of the data-points around
a {\it test model} display a $\overline{\delta}$ that is compatible
with the {\em best-fit} model.
$P\simeq 1$ indicates that the model is a good representation of the
observed spectrum.
 
The left panel of Fig.~\ref{IRtest} shows the average results for the 
observed J and H band spectra of M82. 
It can be easily appreciated that the best-fit solution provides
in all cases a clear maximum in $P$ ($>$99\%)
with respect to the {\it test models}.
More relevant, {\it test models} with an abundance
variation $\Delta[X/H]\ge\pm 0.2$~dex
lie at $\simeq 1.5\sigma$ from the best-fit solution,
while {\it test models} with 
$\Delta[X/H]\ge\pm 0.3$~dex lie at $\sigma \ge 2$ from the best-fit solution.

The analysis of the line equivalent widths provide fully consistent results.
The right panel of Fig.~\ref{IRtest} shows the average  
difference between the model and the observed equivalent width measurements.
Models with $\pm 0.2$~dex abundance variations from the best-fit solution are still 
acceptable at a $\simeq 1.5\sigma$ significance level, while those with $\pm 0.3$~dex 
variations are only marginally acceptable at a $2-3 \sigma$ level.

Models with stellar parameters varying by $\Delta $T$_{\rm eff}$$\pm$200~K, 
$\Delta $log~g=$\pm$0.5 and $\Delta \xi$$\mp$0.5~~km~s$^{-1}$ and abundances 
varying accordingly by 0.1-0.2 dex,   
in order to still reproduce the deepness of the 
observed features, are also less statistical significant 
(on average only at $\ge2 \sigma$ level) with respect to the best-fit solution.    
Hence, as a conservative estimate of the systematic error in the derived best-fit abundances, 
due to the residual uncertainty in the adopted stellar parameters, one can 
assume a value of $\le \pm 0.1$~dex. 

By taking into account the overall uncertainty in the definition of the 
average population and the statistical significance of our spectral synthesis 
procedure, we can safely conclude that the stellar abundances 
can be constrained well within $\pm$0.3~dex and their abundance ratios 
down to $\simeq$0.2~dex, 
since some (if not all) of the stellar parameter degeneracy is removed.

\section{X-ray spectra and hot gas-phase abundances}
\label{xray} 

The determination of the abundances of the hot X-ray emitting gas
in SB galaxies has traditionally suffered from large uncertainties.
Indeed, the low angular and spectral resolution of the various X-ray
telescopes in the pre-\xmm/{\it Chandra} era did not allow to disentangle
between point sources and hot gas emission, making abundance determinations
severely model-dependent \citep{dah00}. The problem lies in
the fact that the X--ray spectrum of SB galaxies
contains two major components: the emission from hot diffuse
gas and the integrated contribution of point sources. The first is described by 
an optically thin thermal spectrum {\it plus} emission lines, 
the latter by a power-law.  
If the angular resolution is not good enough, 
it is not possible to reliably subtract
the point sources from the total spectrum, so that the equivalent
widths of the emission lines (and thus the element abundances) are
not unambiguously defined.

On the other hand, high angular resolution alone is not enough: recent
{\it Chandra } studies (\citealt{stric02}; \citealt{mart02})
demonstrated that the relatively low spectral
resolution of the {\tt ACIS} detector makes individual element abundance
analysis still problematic. The high spectral resolution of the 
{\it Chandra} gratings rapidly degrades for sources which are much more 
extended than the instrumental {\it Point Spread Function} (PSF) and thus 
these gratings are almost useless 
for the study of nearby SB galaxies.
However, by coupling the high angular
resolution ($0.5^{\arcsec}$ PSF) of {\it Chandra} with the high spectral
resolution ($\lambda/\Delta\lambda\simeq 400$) 
of the XMM/RGS, which is not sensitive to 
the source extension, 
it is possible to 
overcome the above described difficulties.

M82 was observed several times by \chandra; we consider the two
longest exposures, of 33 ks and 15 ks respectively, both obtained on
September 1999. A cumulative spectrum of the brightest point sources
was extracted from the two \chandra\ {\tt ACIS-I} observations; the
best-fit model was an absorbed power-law with
$\rm N_{{H}}=7.9\pm0.7\times 10^{21}$ cm$^{-2}$,
$\Gamma=0.84\pm0.07$, $\chi_{{\rm r}}^{2}=0.4$, which is typical of
High Mass X--ray Binaries \citep[see][and reference therein]{pr02}.
The 0.5--10 keV flux was $4.2\times 10^{-12}$ erg s$^{-1}$ cm$^{-2}$,
which represented $\sim25\%$ of the total (point source {\it plus} diffuse)
flux (but notice that M82 is a variable source, \citealt{pg99},
\citealt{rg02}).

\begin{figure}[htbp] 
\plotone{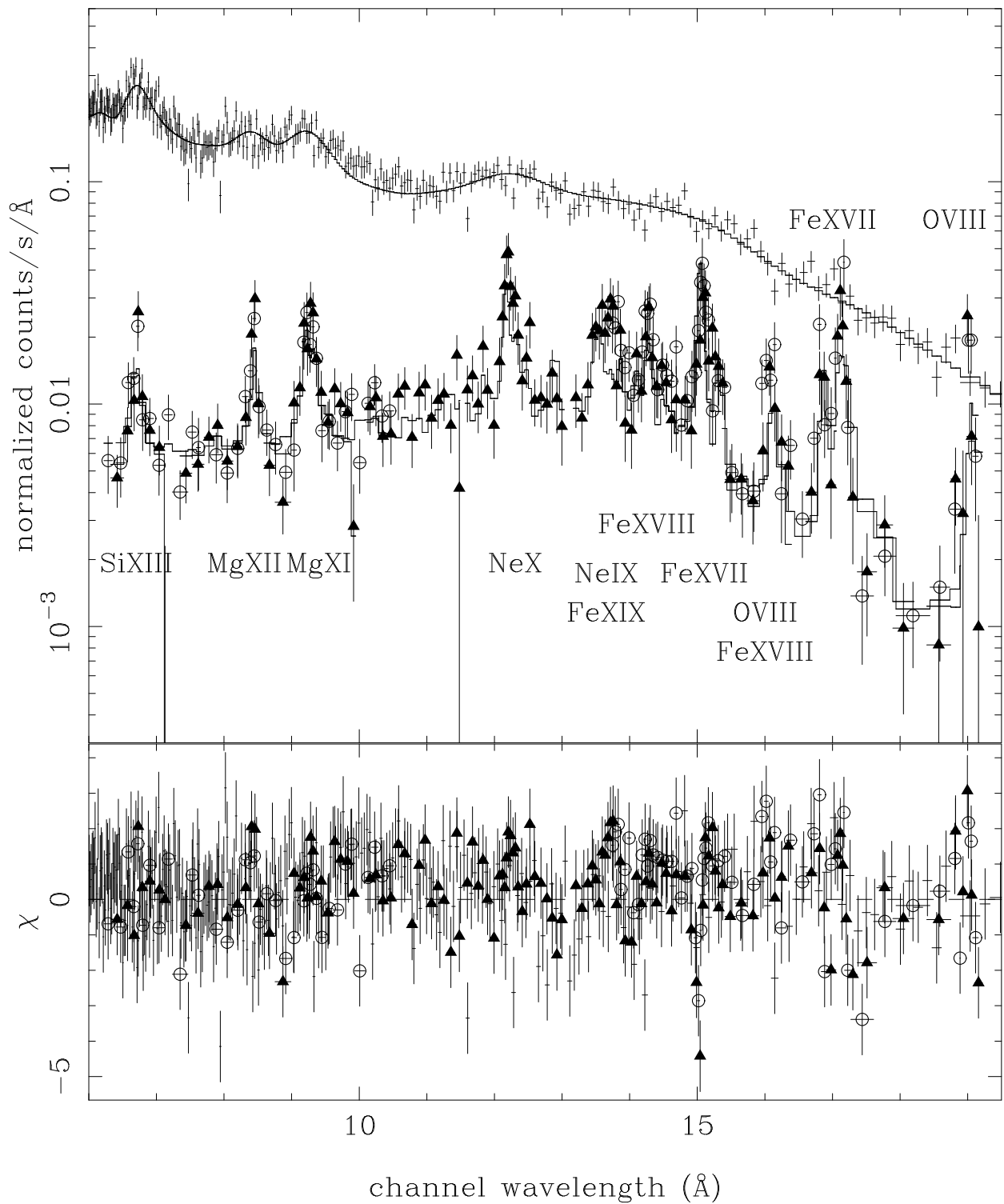} 
\caption{
XMM-RGS spectrum of the nuclear region of M82.
Upper panel: data points 
(lower spectrum - open circles: RGS1; filled triangles: RGS2; upper spectrum: {\it pn})
along with the best-fit model. Lower
panel: residuals in units of $\sigma$.
\label{XR}}
\end{figure}

M82 was observed by \xmm\ for 30 ks in May 2001; in the \xmm\
observation, after light-curve cleaning, $\simeq 20$~ks of data were
available for scientific analysis. We extracted the {\tt EPIC}--{\it pn}
spectrum from a circular region centered on the starbursting core of
M82 and with a $15\arcsec$ radius, which approximately matches the
RGS PSF; the standard recipe for {\tt EPIC} spectra extraction
suggested in the XMM web pages was followed.  The {\tt XMMSAS}
software version 5.3.0 was used.  A background spectrum was taken from
standard background files provided by the \xmm\ SOC. RGS spectra
were extracted with standard settings (i.e.\ including events within
90\% confidence and rejecting as background events those outside 95\% of the PSF
figure); this procedure is similar to the one used by \citet{rs02}\ but
with more recent calibrations. After visual inspection, we discarded
the 2nd-order spectra because of low signal/noise ratio. We included in
the fit only the 1st-order channels comprised in the 6--20 \AA\
interval, since outside the S/N ratio rapidly drops. A
systematic error up to 10\% in the 14--18 \AA\ channels was included
(XMM calibration document XMM-SOC-CAL-TN-0030 Issue 2).

The hard--band (2--7 keV) \chandra\ images show that the point sources emission
largely exceeds that of the hot gas, while 
the hard part (3--7 keV) of the {\em pn} spectrum 
can be well described by a power-law whose best-fit slope resulted in
the same value previously determined for the \chandra\ point sources.
Thus, making use of information from both \chandra\ and \xmm\ observations,
we have been able to constrain  the point source slope, 
absorption and normalization at the
moment of the \xmm\ observation.

We jointly fitted the {\it pn} and RGS spectra with 
{\tt XSPEC Version~11.2} (see Fig.~\ref{XR}) with a two component 
model which is the sum of: \emph{i)} an absorbed power-law (accounting
for point sources) with the slope and absorption parameters fixed
at the best fit values obtained with \chandra\ while normalization was
left free to account for variability; \emph{ii)} an absorbed, optically--thin
thermal plasma emission, with variable line intensities and differential
emission measure (DEM) distribution.
The {\tt DEM} distribution was
modeled with a 6$^{\textrm{th}}$ order Chebyshev polynomial (\texttt{c6pvmkl}
model in XSPEC); with the best-fit parameters it resembles a bell-shaped
curve with a peak at $kT\sim0.7$ keV and FWHM $\sim0.65$ keV (i.e.\ the
plasma temperatures range from $\sim$0.35 keV to $\sim$1 keV), in
agreement with \citet{rs02}. The thermal component is 
absorbed by cold gas with best-fit column density $N_H \simeq 3.8 \times 10^{21}$ 
cm$^{-2}$.

\begin{figure}[htbp] 
\plotone{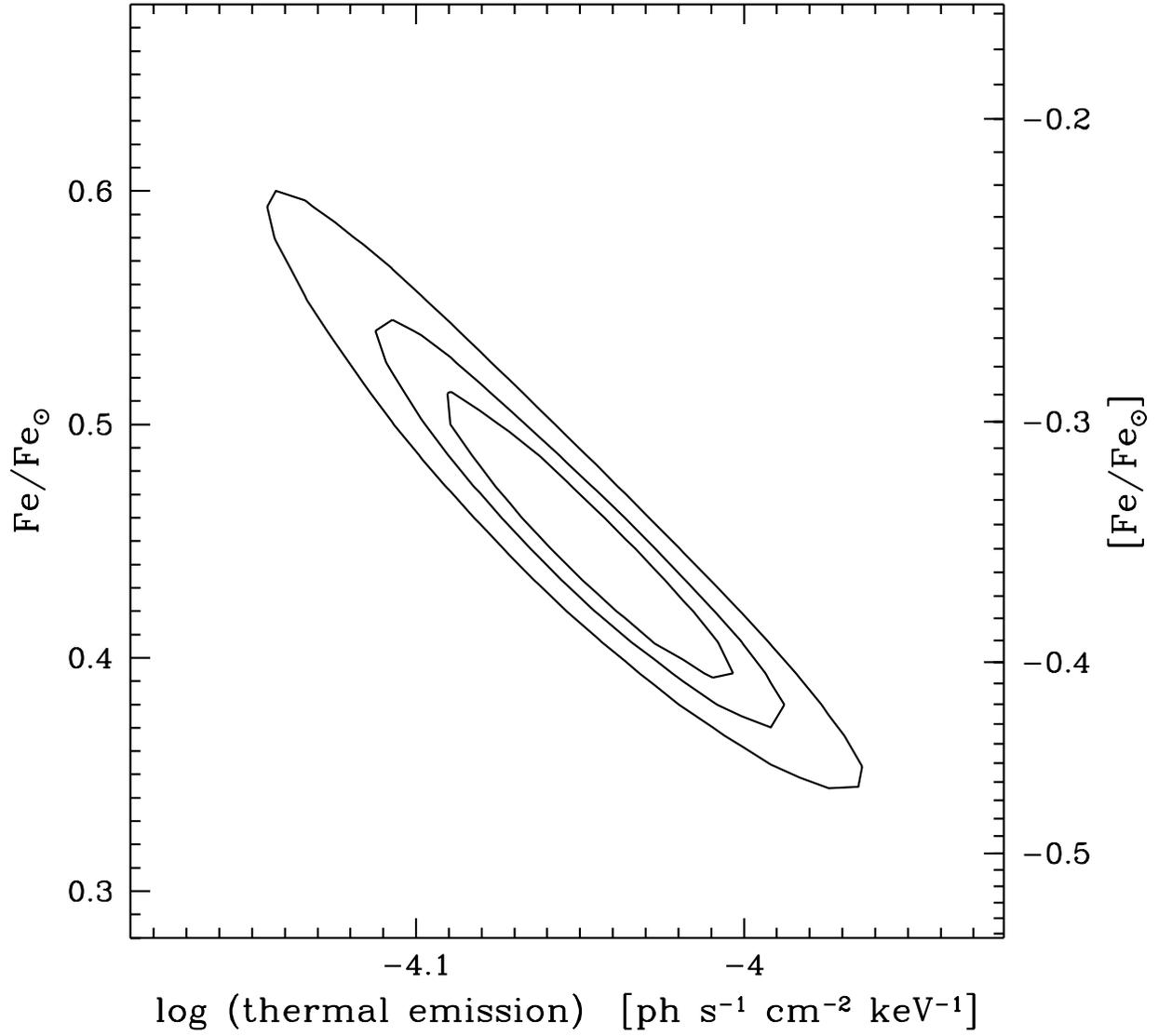}
\caption {68. 90 and 99\% confidence contours of the 
Iron abundance {\it vs} intensity of the thermal component.
For easy of reading  
the y-axis is shown both in linear and logarithmic scale. 
\label{fig-fevsnorm}}
\end{figure}

The confidence contours of the Iron abundance versus normalization of the 
thermal component are reported in Fig.~\ref{fig-fevsnorm}. As expected the
two parameters are well correlated as the higher is the normalization 
of the continuum spectrum the lower is the intensity of the emission lines
and in turn the element abundances.
The derived abundances and their associated 90\% confidence error are
shown in Table~\ref{ab}. We notice that these abundances are about a factor
2 below those reported in \citet{rs02}; we attribute the discrepancy
to the different modeling of the continuum emission (\citealt{rs02}\ did
not include {\it pn} data, neither considered the contribution from point
sources).  However, it is worth of mentioning that both sets of abundances indicate 
a low O abundance compared to the other $\alpha$-elements.

\begin{figure}[htbp] 
\plotone{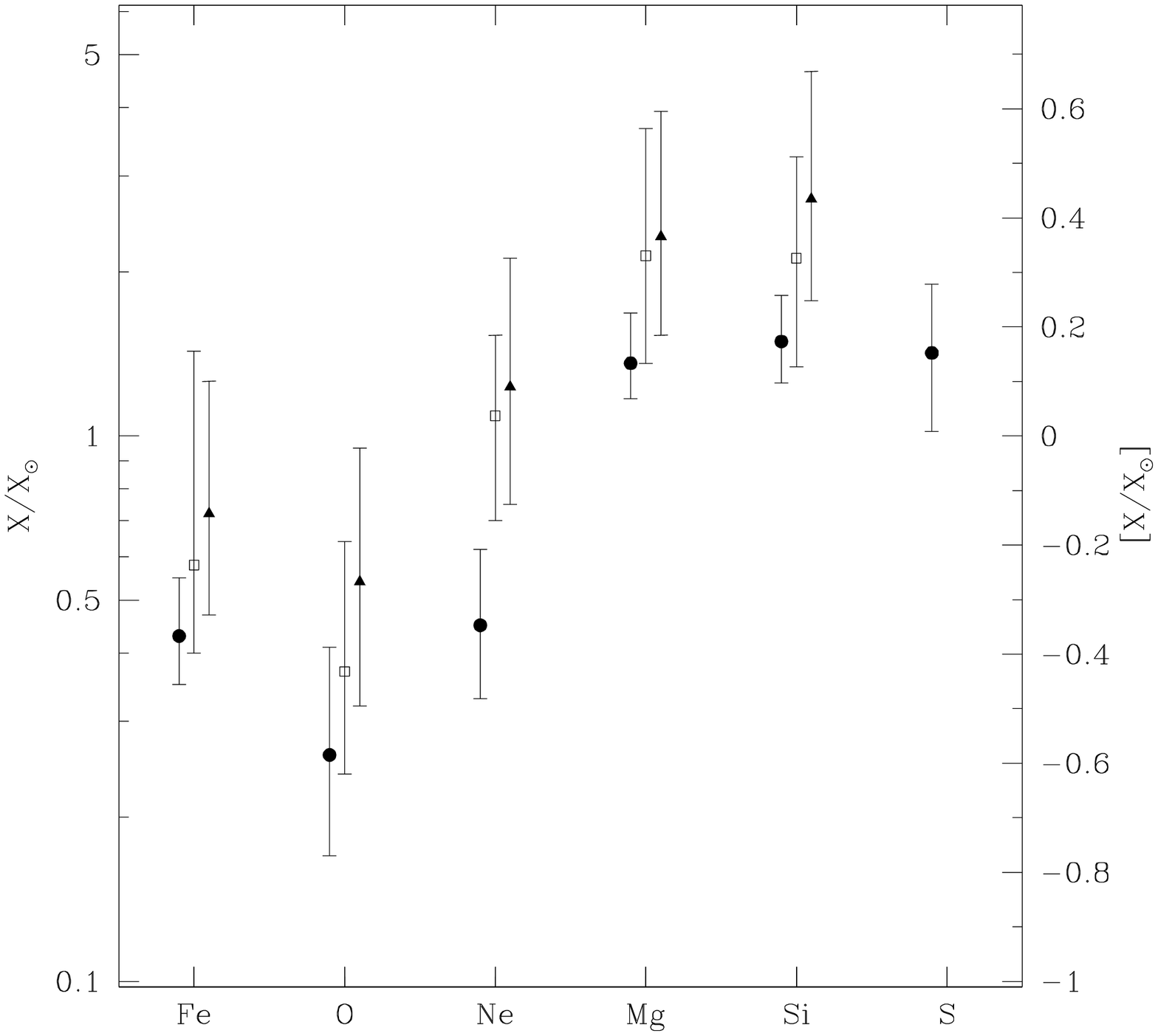} 
\caption {Best-fit element abundances of the hot gas for different models. 
Filled circles: fit 
of both RGS and {\it pn} data, thermal {\it plus} power-law model. 
Open squares: RGS data alone, thermal {\it plus} power-law model. 
Filled triangles: RGS data alone, thermal model alone.
For easy of reading the y-axis is shown both in linear and logarithmic scale. 
\label{fig-abun}}
\end{figure}

To further check the consistency and robustness of our estimates,
we repeated the fit two times considering only the RGS
data and using \emph{i)} the above described 
thermal {\it plus} power-law model and \emph{ii)} a thermal model alone. 
This last check, in particular, 
most closely matches the analysis previously made by
\citet{rs02} to the point that the discrepancies are within the errors
and can be attributed to the differences in the used instrument calibration.
The abundances derived from the three different models
are plotted in Fig.~\ref{fig-abun}. Those obtained combining 
{\it pn} and RGS have significantly smaller uncertainties. 
One can see that different models have mainly 
the effect to almost rigidly scale all the abundances, with minor impact 
on the abundance ratios.  

\begin{figure}[htbp] 
\plotone{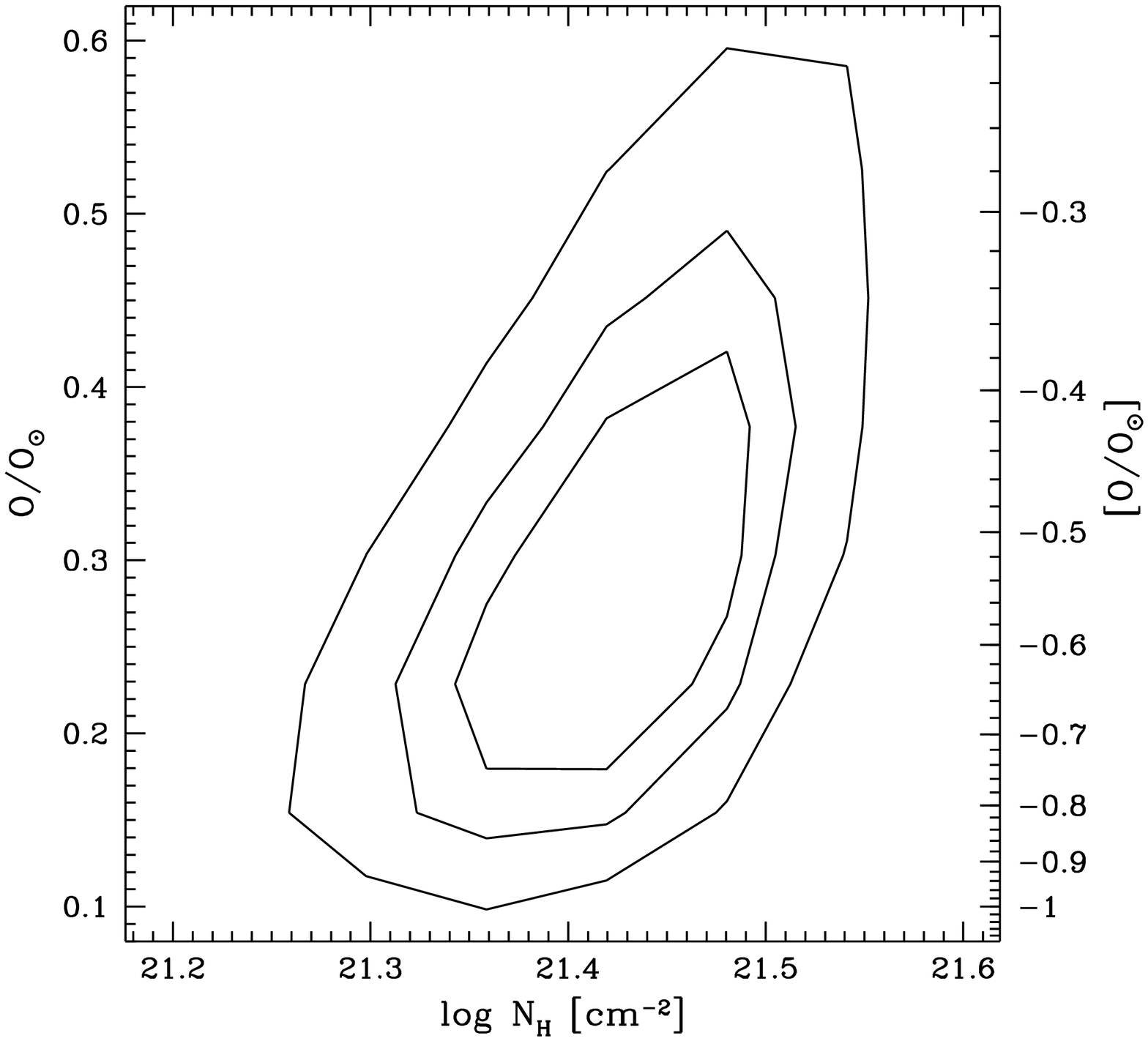} 
\caption {Hot gas-phase Oxygen abundance {\it vs} absorption. 
For easy of reading  
the y-axis is shown both in linear and logarithmic scale. 
The correlation, if any, is
rather weak, and the Oxygen abundance can be well constrained.
\label{fig-ovsnh}}
\end{figure}

The low Oxygen abundance is somehow difficult to explain in the framework
of $\alpha$-element enhancement by type II SN explosions. Thus,
to check its reliability we included in the fit also the channels
corresponding to the 20--23 \AA\ interval, where the O\textsc{VII}
triplet is observed (although with very few counts). We find no significant
difference neither in Oxygen abundance nor in {\tt DEM} 
distribution (the O\textsc{VIII}
$\lambda$19/O\textsc{VII} $\lambda$22 ratio is extremely sensitive 
to the plasma temperature). Ne is also somewhat under-abundant. We checked
for possible instrumental effects, such as variations in the RGS effective
area in proximity of the Ne and O lines; however the effective area
in this region is a rather smooth function of the wavelength, so that
there should be no instrumental issues and these low abundances are likely
to be real. 

Since M82 is observed edge---on, the X-ray emission of its nuclear 
region is subject to quite heavy absorption which could
affect the O lines detected at low energies.
However, as shown in Fig.~\ref{fig-ovsnh}, the contour plot of the Oxygen abundance 
towards the foreground absorption, suggest a rather weak correlation.
In the next section
we discuss other physical possibilities to explain the O under-abundance
issue.

\begin{figure}[htbp] 
\plotone{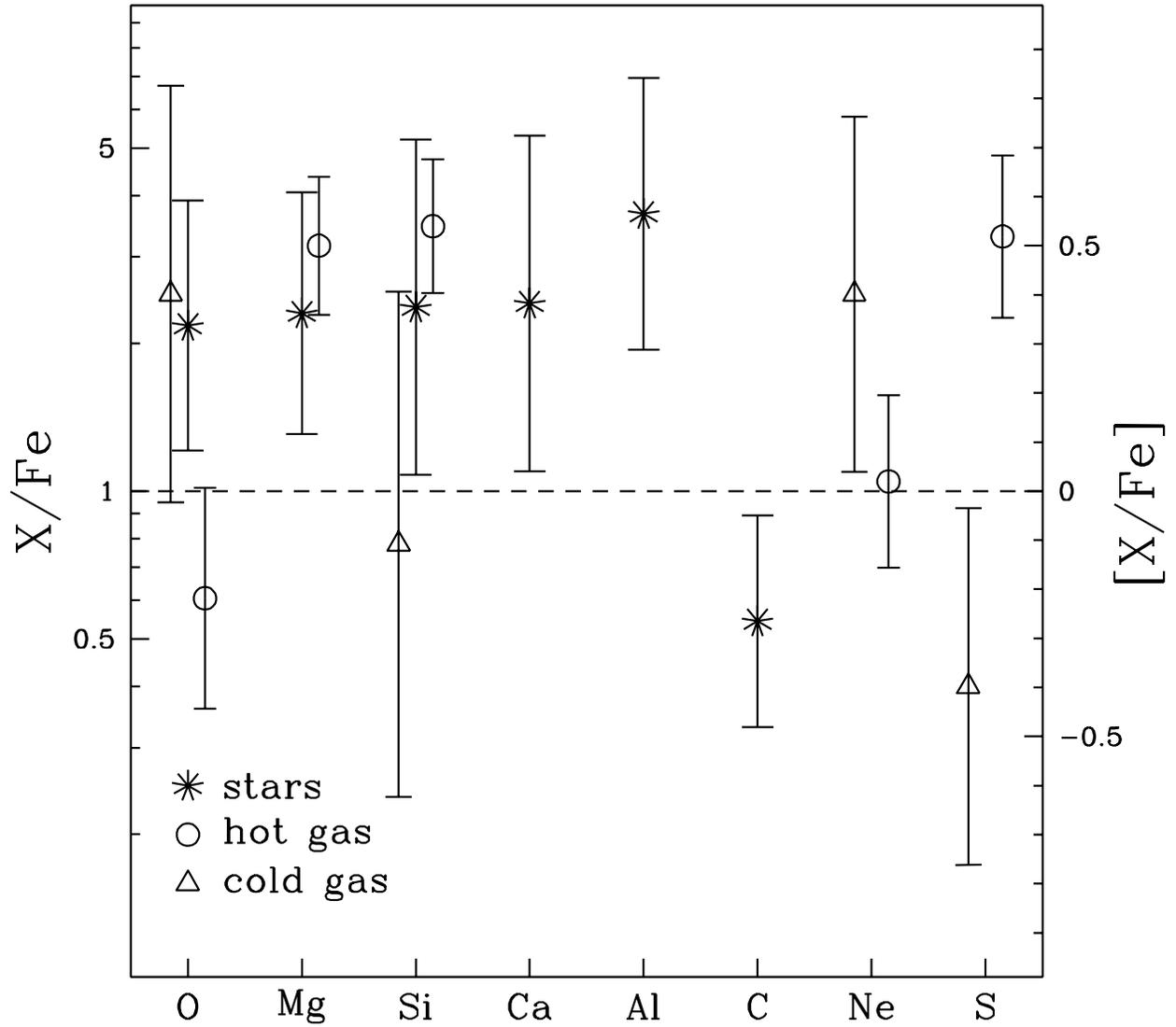} 
\caption {Stellar, hot and cold gas abundance ratios relative to Iron.  
Cold gas-phase abundances are from different data sets in the literature, 
hence they are not homogeneously determined.
The dashed line indicates the solar values.
For easy of reading  
the y-axis is shown both in linear and logarithmic scale. 
\label{abtot}}
\end{figure}

\section{Discussion}
\label{disc}

In M82 all the three components, namely hot and cold gas-phases and stars 
trace a very similar Iron abundance, the average value being 
[Fe/H]$\simeq -0.35$~dex.
Indeed, since Iron is mainly produced  
by SNe~Ia, it is expected to be released in the ISM only after $\simeq 1$~Gyr 
from the local onset of SF. 

At variance, $\alpha -$elements (O, Ne, Mg, Si, S, Ca, Ti)
are predominantly released by SNe~II
with massive progenitors on much shorter timescales.\\
Stars trace an average [$<$Si,Mg,Ca$>$/Fe]$\simeq 0.4$~dex, while the hot gas
suggests an average [$<$Si,Mg,S$>$/Fe]$\simeq 0.5$~dex. 
Such an overall [$\alpha$/Fe] enhancement
in M82 is fully consistent with a standard 
chemical evolution scenario, 
where the ISM in the nuclear regions of massive SB galaxies 
is mainly enriched by the products of SNe~II explosions 
(see e.g.\ \citealt{arn95} for a review) 
occurring in recursive bursts of SF of relatively short duration.

Fig.~\ref{abtot} shows the various abundance ratios relative to Iron, as measured in 
stars, hot and cold gas-phases.
Since the cold gas-phase abundances are from different data sets in the literature, 
hence not homogeneously determined, the inferred abundance ratios should be 
regarded with caution.

When comparing the hot and cold gas-phase abundances, one finds Ne
over-abundant, Si and especially S significantly under-abundant in the
cold gas.  \citet{fs01} suggests S and to a lower level Si
depletion onto interstellar dust grains.  However, other metals like
Fe in particular, with even higher degree of incorporation into dust
\citep{ss96}, should be severely depleted in the cold gas, which does
not seem the case.  A possible explanation is that the optical
lines used to infer Fe and O abundances \citep{acsj79} and the the 
mid/far IR lines used to infer Ne and S abundances \citep{l96,fs01} trace
different nebular sub-structures within the central few hundreds pc,
somewhat chemically dishomogeneous or with different dust content.

The [O/Fe] abundance ratio in M82 is even more puzzling to interpret.
Before the onset of the current burst of SF,
the ISM was enriched in O as well as in the other $\alpha$-elements, 
as suggested by our stellar and cold gas-phase abundances.
The O under-abundance measured in the hot gas 
cannot be modeled with a standard nucleosynthesis 
from the present generation of SNe~II.
\citet{ume02} suggested explosive nucleosynthesis in core-collapse hypernovae 
but they predicted a large under-abundance of Ne and Mg as well, which is 
not observed.   
[O/Fe] under-abundance can be also explained 
with a major Fe enrichment by current SN-Ia explosions 
from previous generations of stars, 
but in this case all the [$\alpha$/Fe] ratios should be low, 
again contrary to what observed.
O might be also locked into dust grains. 
Indeed, SNe~II are known to produce significant amounts of dust 
\citep{tf01} and O, being a basic constituent 
of dust grains, is expected to suffer of severe dust 
depletion. 
However, other metals like Fe in particular, but also Si and Mg 
should be depleted into dust \citep{ss96}, 
leaving almost unchanged or eventually further enhancing the 
[$\alpha$/Fe] ratio. 

\section{Conclusions}
\label{conc}

Our abundance analysis of the stellar and hot gas-phase components 
in the nuclear region of M82 
indicate an Iron abundance about half Solar and an overall 
$\alpha-$enhancement by a factor 
between 2 and 3 with respect to Iron.

These abundance patterns can be easily explained within a standard nucleosynthesis 
scenario where the ISM is mainly enriched by SNe~II on relative short timescales and 
with a star formation process occurring in recursive bursts.  

Oxygen behaves in a strange fashion. It is over-abundant in stars and cold gas, 
similarly to the other $\alpha$-elements, while is significantly under-abundant 
in the hot gas. Major calibrations and/or modeling problems seem unlikely.
Hypernovae nucleosynthesis, dust depletion, SN-Ia enrichment can somehow explain 
an Oxygen under-abundance but other metals should follow the same pattern, while
they do not. 
   
Somewhat exotic threshold effects and/or depletion mechanisms preferentially 
affecting Oxygen could be at work in the nuclear region of M82, but presently 
this issue remains controversial.

\acknowledgments 
The financial support by the Agenzia Spa\-zia\-le Ita\-lia\-na (ASI) 
is kindly acknowledged.

\end{document}